\documentclass[prapplied,rsi,amsmath,amssymb,reprint,superscriptaddress]{revtex4-1}

\usepackage{longtable}

\usepackage{graphicx}
\usepackage{color}

\usepackage[version=3]{mhchem} 
\usepackage{blindtext}
\usepackage{siunitx}
\usepackage{float}
\usepackage[english]{babel}

\begin{document}
\title{Exploring the formation of gold/silver nanoalloys \\ with gas-phase synthesis and machine-learning assisted simulations} 

\author{Quentin Gromoff}
\affiliation{CEMES, CNRS and Universit\'e de Toulouse, 29 rue Jeanne Marvig, 31055 Toulouse Cedex, France}
\author{Patrizio Benzo}
\affiliation{CEMES, CNRS and Universit\'e de Toulouse, 29 rue Jeanne Marvig, 31055 Toulouse Cedex, France}
\author{Wissam A. Saidi}
\affiliation{National Energy Technology Laboratory, United States Department of Energy, Pittsburgh, PA 15236, USA}
\affiliation{Department of Mechanical Engineering and Materials Science, University of Pittsburgh, Pittsburgh, PA 15261, USA}
\author{Christopher M. Andolina}
\affiliation{National Energy Technology Laboratory, United States Department of Energy, Pittsburgh, PA 15236, USA}
\affiliation{Department of Mechanical Engineering and Materials Science, University of Pittsburgh, Pittsburgh, PA 15261, USA}
\author{Marie-José Casanove}
\affiliation{CEMES, CNRS and Universit\'e de Toulouse, 29 rue Jeanne Marvig, 31055 Toulouse Cedex, France}
\author{Teresa Hungria}
\affiliation{Centre de MicroCaractérisation Raimond Castaing, Université de Toulouse, 3 rue Caroline Aigle, F-31400 Toulouse, France}
\author{Sophie Barre}
\affiliation{CEMES, CNRS and Universit\'e de Toulouse, 29 rue Jeanne Marvig, 31055 Toulouse Cedex, France}
\author{Magali Benoit}
\affiliation{CEMES, CNRS and Universit\'e de Toulouse, 29 rue Jeanne Marvig, 31055 Toulouse Cedex, France}
\author{Julien Lam}
\email{julien.lam@cnrs.fr}
\affiliation{CEMES, CNRS and Universit\'e de Toulouse, 29 rue Jeanne Marvig, 31055 Toulouse Cedex, France}
\affiliation{Univ. Lille, CNRS, INRA, ENSCL, UMR 8207, UMET, Unité Matériaux et Transformations, F 59000 Lille, France}

\begin{abstract}

While nanoalloys are of paramount scientific and practical interests, the main processes leading to their formation are still poorly understood. Key structural features in the alloy systems, including crystal phase, chemical ordering, and morphology, are challenging to control at the nanoscale, making it difficult to transfer their usage to industrial applications. In this contribution, we focus on the gold/silver system that has two of the most prevalent noble metals, and combine experiments with simulations to uncover the formation mechanisms at the atomic-level. Nanoparticles are produced using state-of-the-art inert-gas aggregation source and analyzed using transmission electron microscopy and energy-dispersive x-ray spectroscopy. Machine-learning-assisted molecular dynamics simulations are employed to model the crystallization process from liquid droplets to nanocrystals.  Our study finds a preponderance of nanoparticles with five-fold symmetric morphology, including icosahedron and decahedron which is consistent with previous results on mono-metallic nanoparticles. However, we observe that gold atoms, rather than silver atoms, segregate at the surface of the obtained nanoparticles for all the considered alloy compositions. These segregation tendencies are in contrast to previous studies and have consequences on the crystallization dynamics and the subsequent crystal ordering. We finally show that the underpinnings of this surprising segregation dynamics is due to charge transfer and electrostatic interactions rather than surface energy considerations.

\end{abstract}

\maketitle

\section{Introduction}

By combining two or more metallic elements within the same nanoparticle, synergistic properties can emerge and result into innovative technological applications.\cite{Nanoalloys,Calvo2020,Ferrando2016Aug,Ferrando2008Mar}  Numerous research fields including optics, catalysis, bio-medicine and electronics are already considering these so-called nanoalloys mainly because they exhibit the fundamental advantage  of an extremely rich structural landscape with a variety of shapes, chemical orderings and crystalline phases. However, since the physical and the chemical properties of nanoalloys are intrinsically related to their internal structure, advances in this field are strictly constrained by synthesis experiments, and it becomes crucial to better understand the intricate relationship between the experimental conditions and the obtained structures. Addressing this pivotal challenge first comes directly from experimental studies, where two types of complementary approaches are usually considered: (1) Systematic variations of experimental conditions are followed by post-mortem structural analysis\cite{Penuelas2008Mar,Amendola2014Jan,Warkocka2015May,Ahmad2019Oct,Benzo2019Sep,Wang2022Oct} and (2) In situ experiments allow for a direct observation of the nanoalloys formation.\cite{vanderHoeven2018Aug,Liu2013Dec,Carrillo2018Aug,Ramade2017Sep,Mertens2011Sep,Chen2022Apr,Ahmad2018Feb} 

Computational simulations have been the ideal complementary tool because it offers an unambiguous atomistic picture and enables standardized examinations of various experimental parameters. In this context, the dynamics of nanoalloy formation are usually investigated with classical force fields via Monte Carlo and molecular dynamics simulations, typically employed to study large-scale systems.\cite{Goudeli2017Nov,Nelli2020Jul,Nelli2019Jul,Forster2019Oct,Hizi2023Jan}  The derived predictions, however, often lack chemical accuracy and can hardly be used to draw quantitative conclusions. As an alternative,  quantum simulations based on first-principles density functional theory (DFT) have been performed in order to address this accuracy issue.\cite{Chen2008Jan,Demiroglu2016Sep,Barcaro2011Apr,Tan2012Sep,Davis2014Jan} However, due to their high computational costs, these methods are often limited to the equilibrium properties of bulk systems or small clusters. Recently, machine-learning approaches have been proposed to bridge the gap between these two approaches. Indeed, machine-learning interaction potentials (MLIP) are constructed by combining a very complex mathematical formulation with numerous fitting parameters along with an extensive DFT-generated database composed of different structures in conjunction with their associated energies, forces, and virials. Evidence of the success of these methods can be seen through the very diverse types of materials that were so far modeled with MLIP, including metals, oxides, carbon and silicon-based, organics, and perovskites.\cite{Wisesa2022-oxides-ML,Wang2022-perovskites-ML,Andolina2020-CuZr-ML,Andolina2021-AlMg-ML,Bayerl2022-ML-conv,Erhard2022Apr,Mocanu2018Sep,Gartner2022Dec,Piaggi2022Aug,Panizon2015Oct,Benoit2020Dec,Tallec2023Jan,Laurens2021Jun,Goniakowski2022Oct,Lam2020Mar,Caro2018Apr,Deringer2017Mar} Prompted by these advances, DFT accurate large scale simulations can finally be carried out to investigate the intricate formation processes occurring in nanoparticle synthesis. 

Herein we studied the formation of AuAg nanoalloys that have been considered in many applications owing to, in particular, their plasmonic,\cite{Tsao2014Jan,daSilva2016Jun,Newmai2022Feb} catalytic\cite{Tsao2014Jan,Newmai2022Feb,Han2014Jun,Urzua2022} and antibacterial\cite{Li2018Aug,Velmurugan2022Mar,Padmos2015Apr,Ding2017Jan} properties. In this context, targeted technological applications require the control over two principle structural parameters. First, different morphologies can be stabilized with a competition between truncated octahedron structures and five-fold symmetric morphology including icosahedron and decahedron that originates from the balance between cohesive, surface and  elastic strain energies.\cite{Barnard2009Jun,Wang2012Jun,Pohl2012Dec,Liao2018Apr,Belic2011Oct,Casillas2012Apr,Garden2018Mar,Wang2011Jun,Myshlyavtsev2017Jul,Baletto2001Mar,Baletto2002Mar} Second, while metallic species can be found in different chemical arrangements with the possibility of surface segregation, there is currently no consensus in the literature on whether gold or silver is more likely to segregate to the surface. \cite{Yen2009Oct,Liao2020Nov,Guisbiers2016Jan,Stein2022Aug,Liao2018Apr,Martinez2012Jul,He2018Jan,Du2019Aug,Bon2019Aug,Paz-Borbon2008Apr,Deng2011Jun,Cerbelaud2011May,Chen2008Jan,Chen2008Jun,Hoppe2017Dec,Rapetti2019Mar,Wang2021Dec,Gould2014Sep}

In this contribution, we experimentally show that gas-phase synthesis can lead to decahedral and icosahedral AuAg nanoalloys both displaying unambiguous gold surface segregation. Our machine-learning assisted simulations confirmed those experimental findings and enabled investigations over a wider spectrum of chemical compositions. Moreover, the simulations allow us to go beyond post-mortem analysis, thus uncovering how gold segregation can affect the nucleation process triggering the nanoparticle crystallization at the atomistic level. When compared to the current literature, the novelty of this work is three-fold: (1)\,Pentatwinned decahedra/icosahedra are not only present in mono-metallic systems and can also be stabilized in the case of Ag/Au nanoalloys, (2)\, Gold atoms can segregate at the nanoparticle's surface even with equimolar and silver-rich compositions and (3)\,Machine-learning assisted simulations can be used to model nanoalloys accurately  to observe atomic scale processes occurring during their formation.

\section{Methodology}

Two complementary approaches were applied to investigate the formation of AgAu nanoalloys. We used gas-aggregation magnetron-sputter deposition from two elemental targets of gold and silver to synthesize the AuAg nanoparticles. Note that this synthesis process involves the nucleation and growth of the nanoparticles inside a gas aggregation chamber, before their landing on the substrate\cite{Huttel2017Mar}. These nanoparticles are extremely pure, i.e., free from any ligand or surfactant. The synthesized nanoparticles were further investigated using high-resolution transmission electron microscopy (HRTEM) and high-angle annular dark-field scanning TEM (HAADF-STEM) for uncovering their structural and morphological details. The chemical distribution of the two elements inside a given nanoparticle was analyzed by energy-dispersive x-ray spectroscopy (EDS-STEM). In concert, machine-learning-assisted simulations were carried out. In particular, we used a deep neural network potential (DNP) that was previously developed by Andolina et al.\cite{Andolina2021Aug} and was further tested under numerous conditions\,[see SI\,A]. A key advantage of using the DNP approach is that, although this is not yet sufficient for reaching the observed experimental sizes ($\approx$7.5\,nm),  we still managed to perform molecular dynamics (MD) with a much longer duration and a larger number of atoms ($\approx$15000 atoms ie. 6.3\,nm) than can be expected from typical DFT calculations.


\subsection{Experimental setup}

AuAg nanoparticles were grown in a water-cooled Nanogen-Trio Gas Aggregation Source (GAS), from Mantis Deposition Ltd. The nucleation and growth of the particles are obtained through DC magnetron co-sputtering of extremely pure (99.99\%) Au and Ag targets of diameter equal to $1$ inch, located side by side on an integrated magnetron sputtering head, positioned at 90 mm from the exit slit of the aggregation zone. The DC magnetron current, applied independently to the two targets, was fixed to 40 mA for both gold and silver. With the synthetic method, it would be possible to control the stoichiometry by using an alloyed target with precise chemical composition. However, studying the influence of the chemical composition would then require  fabrication of targets with different stoichiometry. For each of them, the sputtering property would not necessarily be similar which should change the nanoparticle size distribution. In order to study the influence of the chemical composition for a fixed size distribution, we opted instead to use two elemental targets with a specific set of electrical powers applied in each of them. This allows us to obtain different chemical compositions in a single experiment all with the same sputtering conditions. Before depositing on the ultra-thin-carbon coated copper TEM grid, located in the ultra-high-vacuum deposition Chamber (with a base pressure of $10^{-9}$ mbar), the nanoparticles were size-selected by a Quadrupole Mass Filter installed between the GAS and the deposition chamber. The obtained nanoparticles were studied through high-resolution TEM (HRTEM), atomically resolved high-angle annular-dark-field scanning TEM (HAADF-STEM) and energy-dispersive x-ray spectroscopy (EDS) experiments. A Cs corrected 200 kV FEI Tecnai F20 microscope was used for HRTEM studies, and a probe corrected Jeol ARM200F microscope was used for HAADF-STEM and EDS studies. We note that no oxidation of the nanoparticles was observed even if the EDS analysis was performed several months after the synthesis.

\subsection{Atomistic simulations}

Regarding the interaction potentials, the complexity and non-linearity of deep neural networks allow the development of interaction potentials capable of making predictions close to DFT accuracy. The deep neural network interaction potential for Au-Ag (DNP) developed by Andolina et al.\cite{Andolina2021Aug} using the DeepPot-SE\cite{Zhang2018May} method of the DeePMD-Kit\cite{Wang2018Jul} was selected for this work because of its excellent accuracy with pure as well as alloyed systems [See SI\,A]. All DFT calculations were performed with VASP\cite{KressePRB1993,Kresse1996Jul,Kresse1996Oct} using the Perdew-Burke-Ernzerhof\cite{pbe} (PBE) functional and the projector augmented wave (PAW) method\cite{KressePRB1999} with an energy cutoff of $\SI{400}{eV}$. 

The molecular dynamics (MD) trajectories are obtained using the Large-scale Atomic/Molecular Massively Parallel Simulator (LAMMPS) coupled to the DNP. We employed constant volume and constant temperature (NVT) ensemble in these simulations with a time-step of 1\,fs and a damping coefficient of 100\,fs. Liquid droplets are obtained by melting nanoparticles of three sizes (250, 500 and 750 atoms with respectively 1.4\,nm, 2.0\,nm and 2.4\,nm) for three stoichiometries (Ag$_3$Au, AgAu and AgAu$_3$) at 2000\,K. For the freezing simulations, we utilized a cooling rate of $\SI{2e11}{K/s}$ from $\SI{750}{K}$ to $\SI{350}{K}$. The process is repeated three times with different initial velocities. The employed numerical setup that consists in freezing disordered liquid droplets was already used to simulate nanoparticle formation and mimic gas-phase synthesis\cite{Amodeo2020Oct,Snellman2021Jun,Nelli2023Jan}. In addition, to assess the stability of our obtained systems, we also carried out hybrid MD/MC simulations that consists of 10 Monte Carlo (MC) moves combined with 10 atomic species swapping every 100 MD time steps. 

We note that our choice of simulation protocol assumes that the nanoparticles are formed from the liquid phase. However, we must note that similar experiments have also been modeled with atom by atom growth directly from solid precursors \cite{Goniakowski2010Apr,Wells2015Apr,Xia2021May}. At this stage, it remains difficult experimentally to know which formation pathway is more likely to occur in our experimental protocol.

Finally, simulations are analyzed using Ovito\cite{Stukowski2009Dec} built-in functions.  Common neighbor analysis  is employed to measure crystal ordering \cite{Stukowski2012May}. To obtain the size of the largest ordered cluster, atoms with crystalline order are grouped together in clusters within a cut-off equal to 3.5\,\AA.

\section{Results}

\subsection{Morphology}

\begin{figure*}
    \includegraphics[width = 17cm]{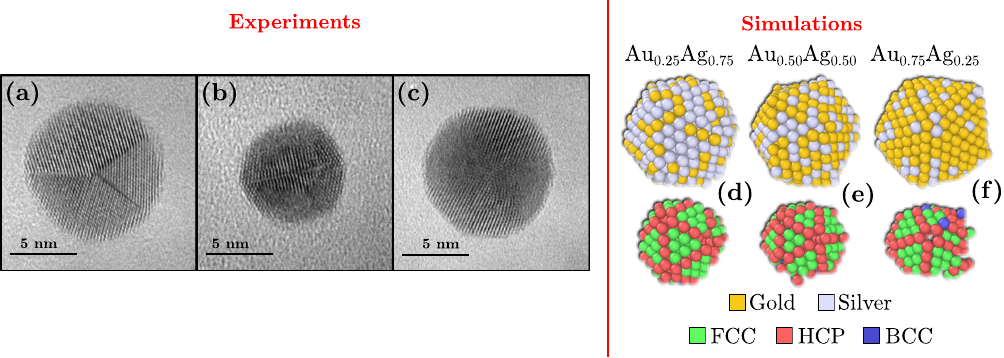}
    \caption{(Left) HRTEM images of 3 representative AuAg nanoparticles. (a) a decahedral NP observed along its five-fold symmetry axis, (b) an icosahedral NP observed along its two-fold symmetry axis, (c) an icosahedral NP observed close to its 3 fold-symmetry axis. (Right) Typical nanoparticles obtained after freezing using machine-learning-assisted MD simulations for three different chemical compositions along with the corresponding crystal analysis as obtained using common neighbor analysis. Non-crystalline atoms (including surface atoms) were removed for clarity.}
    \label{fig:morpho}
\end{figure*}

We begin by analyzing the morphologies of both the experimentally synthesized and the simulated structures. The HAADF-STEM observations show that the synthesized nanoparticles are well-dispersed on the substrate and display a narrow size distribution with a mean diameter of $7.4$ nm and a standard deviation of $0.75$ nm (see SI\,B). When suitably oriented along (or close to) one of their low-index zone axes, the synthesized nanoparticles can be clearly identified as either decahedra or icosahedra,\cite{FD9919200173} both exhibiting triangular surfaces dominated by (111) orientations. Fig.\,\ref{fig:morpho}(a-c) show three different HRTEM images of representative nanoparticles, one $10.5$ nm large decahedron (Fig.\,\ref{fig:morpho}(a)), one $7.5$ nm large icosahedron (Fig.\,\ref{fig:morpho}(b)) and one $9.9$ nm large icosahedron (Fig.\,\ref{fig:morpho}(c)). Similarly, although not with the same size because of the computational cost of the numerical freezing setup, our MD simulations in conjunction with the DNP were able to reproduce the (111) surface preponderance signature of the icosahedron and decahedron shapes\,[See Fig.\,\ref{fig:morpho}(d-f)]. In addition, DNP simulations allow for characterizing the crystal structure at the atomic scale [Fig.\,\ref{fig:morpho}(d-f)]. In all cases, the facets consist of face-centered cubic triangles whose edges are characterized by hexagonal close-packed atoms which is consistent with the (111) orientation of the surfaces. To better characterize the crystallinity of the simulated nanoparticles, we also measured the atomic strain distribution with respect to the radial position [See SI\,E]. The strain map shows overall a very small deviations from the bulk. The latter is slightly more pronounced for atoms located at droplet surface, which can be attributed to an inherent surface relaxations. Importantly, 
out hybrid MD/MC simulations did not observe any structural modification thus confirming the stability of five-fold symmetric shapes [See SI\,C]. Altogether, our results suggest that icosahedron/decahedron can be stabilized instead of truncated octahedron even with in Au/Ag nanoparticles in the investigated size regime,  i.e. up to $10.5$ nm in experiments and up to $2.4$\,nm in simulations.


For metallic systems in general, icosahedron/decahedron and truncated octahedral shapes are considered more stable respectively at small and large sizes according to surface vs. volume energy considerations\cite{Barnard2009Jun,Wang2012Jun,Pohl2012Dec,Casillas2012Apr,Garden2018Mar,Wang2011Jun,Myshlyavtsev2017Jul,Baletto2001Mar,Baletto2002Mar,Pohl2012Dec} 
However, the size threshold allowing for the transition between the two shapes is still highly debated. 
In experiments for mono-metallic systems, icosahedron and decahedron were usually obtained by physical methods of synthesis.\cite{Barnard2009Jun,Wang2012Jun,Pohl2012Dec,Belic2011Oct} Furthermore, previous studies studied gold nanoparticles of diameters up to 10\,nm and showed that the as-obtained icosahedron/decahedron-shaped nanoparticles remain stable after long periods of electron irradiation in the TEM experiments\,\cite{Wang2012Jun,Barnard2009Jun}, demonstrating that, even when obtained under non-equilibrium conditions, both icosahedron and decahedron are thermodynamically more stable than truncated octahedron. So far, such experimental results were only obtained with mono-metallic systems including both silver and gold. Meanwhile, from the simulation viewpoint, the literature regarding this competition in shape also focused solely on mono-metallic systems. In this context, early works based on semi-empirical interactions potentials obtained much smaller crossover sizes leading to truncated octahedral particles being more preponderant even at sizes in the 1 \,nm to 10\,nm regimes\cite{Baletto2001Mar,Baletto2002Mar,Myshlyavtsev2017Jul,Settem2018Jun}. However, most recent results combining DFT accurate models with equilibrium thermodynamic approaches based on Helmholtz free energy demonstrated instead the decahedral stability \cite{Barnard2009Jun,Garden2018Mar} for nanoparticles up to 15\,nm. Similar to previous studies focusing on mono-metallic systems, our study shows with alloyed AuAg nanoparticles that icosahedron/decahedron shapes are stable even for diameters up to $10.5$\,nm. Altogether, our results are therefore consistent with the literature obtained in the mono-metallic regime. Finally, we must note that stability of non-crystalline structures including decahedra can be favored when mixing different metals in a nanoparticle like in AgCu.\cite{Roncaglia2021Oct,Langlois2012May} Furthermore, it can not be ruled out that the observed decahedra may also result from kinetic trapping at the early stages of the growth processes.

%

\subsection{Chemical ordering}

\begin{figure*}[ht]
    \includegraphics[width = 17cm]{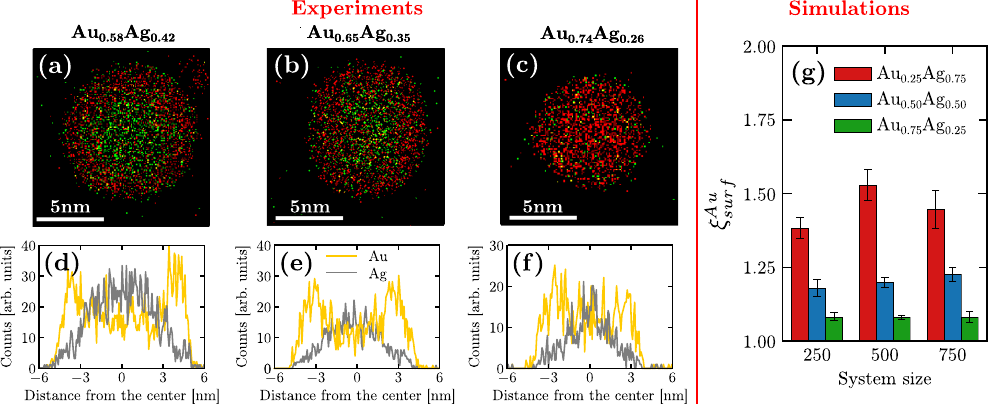}
    \caption{ (Left) (a-c) Experimental EDS measurements for three nanoparticles with different chemical compositions. Au and Ag elements are respectively colored in red and green. (d-f) Distribution of atomic species for the corresponding nanoparticles. (Right) (g) Surface chemical composition obtained in machine-learning assisted MD simulations. Results are shown for different chemical compositions and nanoparticle sizes. Error bars are obtained from averaging over three independent initial conditions.}
    \label{Segregation}	 
\end{figure*}

By using EDS measurements, we were able to investigate the chemical distribution inside a given nanoparticle. Fig.\,\ref{Segregation}(a,b,c) displays the chemical maps of isolated nanoparticles with composition \ce{Au_{0.58}Ag_{0.42}}, \ce{Au_{0.65}Ag_{0.35}} and \ce{Au_{0.74}Ag_{0.26}}. Interestingly, these different compositions were observed in the same sample. Such composition variations can be attributed to a slight evolution of the synthesis conditions such as targets race track and temperature during the deposition time, which was in our case as long as 15 minutes, a duration necessary to collect a sufficiently large number of nanoparticles on the TEM grid in our setting.  We were thus able to analyze nanoparticles with different compositions although silver-rich nanoparticles were not observed experimentally. A striking feature in all these maps is the occurrence of a non-homogeneous distribution with gold atoms segregating at the surface for all of synthesized chemical compositions. For a more quantitative picture,  density profiles recorded along the nanoparticles diameter are presented in Fig.\,\ref{Segregation}(d,e,f). These profiles confirm the gold surface segregation, the silver atoms being mostly confined in the nanoparticle core. In greater details, for nanoparticles with composition \ce{Au_{0.58}Ag_{0.42}}, \ce{Au_{0.65}Ag_{0.35}} and \ce{Au_{0.74}Ag_{0.26}}, we obtained the following respective Au/Ag ratio 1.41, 1.29 and 1.10 by averaging the Au content over the 5 external atomic layers ($\sim$1.25 nm). Note that gold surface segregation is also qualitatively confirmed by HAADF-STEM observations owing to the sensitivity of this technique to the atomic number of the encountered elements (z-contrast) (see SI\,B). 

In order to span a larger range of chemical composition, we complement the experimental observations with numerical simulations.  We separate surface atoms from the bulk based on their lower coordination numbers ($<10$). Fig.\,\ref{Segregation}(g) shows the ratio between the gold proportion at the surface and in the entire nanoparticle denoted $\xi_{surf}^{Au}$. As seen in the figure, our results confirm that gold segregates to the surface even in the regime of silver-rich nanoparticles [see SI\,D for the density profile obtained in simulations]. 

An additional finding is that the gold surface segregation is larger for \ce{Au_{0.25}Ag_{0.75}} than for the less silver-rich nanoparticles. In order to better understand the gold surface segregation obtained at the end of the freezing simulation, we also measured the temporal evolution of the mean square displacement in the initial liquid regime[See Fig. SI\,F]. While all atoms located in the center of the droplet exhibit similar behavior for the mean square displacement, atoms located at the surface possess different diffusion properties depending on their chemical nature or the chemical composition of the droplet. In particular, the diffusion at the surface is lowered when increasing the gold composition. More importantly, silver atoms are always more diffusive than gold atoms at the surface which is consistent with silver atoms being less stable at the surface and showing the tendency to migrate inside the core of the droplet.

\begin{figure}[ht]
    \includegraphics[width = 8.6cm]{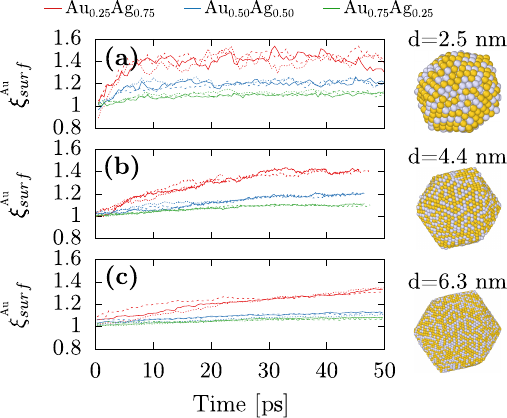}
    \caption{Temporal evolution of the surface chemical composition obtained in machine-learning assisted MD simulations combined with Monte-Carlo and atomic species swap initialized with icosahedron nanoparticles for three different nanoparticles sizes  923, 5083 and 14993 atoms which correspond respectively to 2.5\,nm (a), 4.4,\,nm (b) and 6.3\,nm (c).}
    \label{Segregation2}	 
\end{figure}

Because these first numerical results were obtained from fully freezing liquid droplets, we could only consider systems not larger than 750 atoms which correspond to diameters of $2.4$\,nm. Therefore, we also performed simulations with an alternative protocol to study larger nanoparticles of diameters up to 6.3\,nm. The systems are initialized in the icosahedral structures that circumvent the necessity of having to simulate the whole freezing mechanisms while enabling for directly reaching the previously observed morphology. The atomic chemical species between gold and silver are then randomly assigned to correspond to the three studied stoichiometries of Ag-Au alloys. MD simulations are combined with MC moves and atomic species swapping at 600\,K, which is large enough for atomic swap to operate while maintaining the crystal ordering and overall morphology. By starting with ordered structures yet with chemical disorder, we can focus on the temporal evolution of the surface composition while simulating the larger nanoparticles. In particular, we studied nanoparticles made of 923, 5083 and 14993 atoms, which correspond respectively to 2.5\,nm, 4.4,\,nm and 6.3\,nm. Fig.\ref{Segregation2} shows that convergence is already obtained after 10\,ps for the smallest studied systems while the others would require unreasonably higher computational times. Yet, it remains that in all cases,  while starting at a value of $1$, $\xi_{surf}^{Au}$ monotonically increases with time indicating that similar to the smaller nananparticles gold segregation at the surface exists at the larger sizes,  which are consistent with the experimental results. Moreover, similar to previous results obtained after the freezing simulations for the smaller nanoparticles, the gold surface segregation remains higher in the silver-rich stoichiometries. 

Altogether, both our experimental and our numerical results demonstrate that gold surface segregation can be obtained in Au/Ag nanoparticles regardless of the size and the composition. In experiments, we show that it happens for gold-rich systems while in simulations, we confirm the experimental results and predict that it remains true even for silver-rich systems.

\subsection{Study of the crystallization dynamics}

\begin{figure*}
    \includegraphics[width = 17cm]{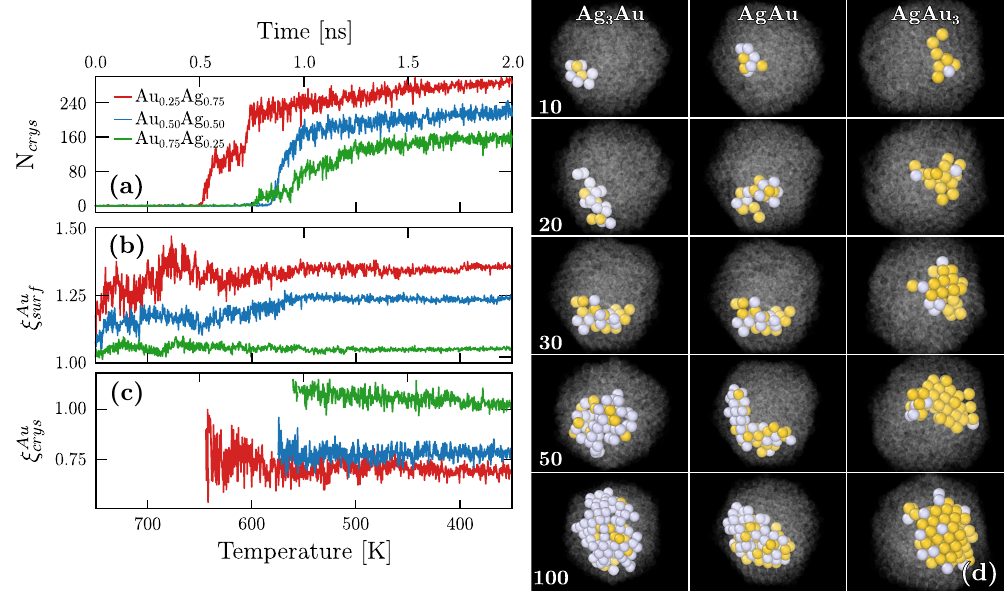}
    \caption{(a-c) Temporal evolution obtained in MD freezing with nanoparticles of 750\,atoms. $N_{crys}$ is the number of atoms within the biggest crystalline cluster. $\xi_{surf}^{Au}$ (resp. $\xi_{crys}^{Au}$) denotes the ratio between the gold proportion at the surface (resp. in the biggest crystalline cluster) and in the entire nanoparticle. We note that for $\xi_{crys}^{Au}$ results are only shown when the crystalline cluster is big enough ie. $N_{crys}>50$. (d) Corresponding images obtained during the crystallization. Non-crystalline atoms are rendered in transparency and grey (resp. yellow) spheres correspond to silver (resp. gold) atoms.}
    \label{fig:temporal}
\end{figure*}

A key advantage of our DNP simulations over experiments, and also DFT calculations that usually employ 0\,K minimization to explore the structural landscape, is that we can retrieve the crystallization dynamics at the atomistic scale. To this end, we follow the evolution of the crystal ordering by measuring the size of the biggest ordered cluster [$N_{crys}$ in Fig.\,\ref{fig:temporal}(a)] and observe that crystallization starts at different temperatures depending on the chemical composition. In particular, silver-rich nanoparticles crystallize first and also reach a higher final value for $N_{crys}$. Fig.\,\ref{fig:temporal}(b) shows the temporal evolution of $\xi_{surf}^{Au}$. While oscillations are observed, $\xi_{surf}^{Au}$ consistently remains greater than one, indicating that gold segregation already occurs in the liquid regime at the early stages of freezing. A plateau in $\xi_{surf}^{Au}$ is also reached after 1\,ns and bellow 550\,K. In the same intermediate times, according to Fig.\,\ref{fig:temporal}(a), the plateau for $N_{crys}$ is not yet fully achieved, suggesting that chemical ordering happens beforehand. We further define $\xi_{crys}^{Au}$ as the relative chemical composition that was previously measured, but within the largest crystalline cluster as opposed to the surface. Fig.\,\ref{fig:temporal}(c) shows that $\xi_{crys}^{Au}$ also quickly converges when compared to $N_{crys}$ thus confirming that chemical ordering occurs first. The value of the observed plateaus are all below or close to $1$, which is consistent with $\xi_{surf}^{Au}$ being above $1$. Indeed, the more gold atoms are present at the surface, the more silver atoms are within the crystalline core of the nanoparticles. For this analysis, while we only showed the temporal evolution of one nanoparticle per chemical composition, SI\,G shows two additional cases that exhibit similar behavior. Finally, the associated snapshots for different sizes of the largest crystalline cluster can be retrieved from those temporal evolutions [see Fig.\,\ref{fig:temporal}(d)]. From these images, one can observe that crystallization emerges first at the periphery of the droplet instead of its core and slowly grows towards the core [See SI\,H for a more quantitative confirmation]. 

Altogether, our findings show that the gold surface segregation occurs already in the liquid regime which has consequences on the subsequent crystallization process. First, we note that gold possesses a larger melting temperature both in bulk and at nanoscale thus indicating that gold crystals are more stable than silver ones. However, in the silver-rich system, the atoms available to trigger crystallization are mostly silver since the majority of gold atoms are at the surface. On the contrary, in the gold rich system, there is a mixture of the remaining gold atoms with all the silver atoms. In one case, a nearly pure metal crystallizes, but in the other, nucleation occurs in an alloying regime which is less favorable. As such, the fact that silver-rich nanoparticles are more crystalline in our simulations is a consequence of the initial chemical ordering that is established already in the liquid regime.

\section{Discussion}

The surface segregation in gold/silver nanoparticles has been the subject of numerous studies with experimental\cite{Yen2009Oct,Liao2020Nov,Guisbiers2016Jan,Stein2022Aug,Martinez2012Jul} and numerical\cite{He2018Jan,Du2019Aug,Bon2019Aug,Deng2011Jun,Cerbelaud2011May,Chen2008Jun,Rapetti2019Mar,Wang2021Dec,Gould2014Sep,Hoppe2017Dec,Paz-Borbon2008Apr,Moreira2023Feb} approaches. Experimentally, chemically-synthesized nanoparticles are found to exhibit silver surface segregation,\cite{Yen2009Oct,Liao2020Nov,Guisbiers2016Jan} which was explained in part by the presence of the oxidized surfaces\,\cite{Hoppe2017Dec}. To the best of our knowledge, only two experimental studies observed surface segregation in Ag-Au nanoalloys made with physical routes of synthesis which is crucial for comparison with our results since they allow for the effects of the ligands and the liquid solvent to be ruled out\,\cite{Martinez2012Jul,Liao2018Apr}. The first study shows surface segregation occurring for the most preponderant chemical species and explained these observations by kinetic trapping at the early stages during growth\cite{Liao2018Apr}. On the one hand, in the gold-rich system, their result is similar to ours although we will show later that kinetic trapping is not the only possible stabilizing effect and that charge transfer can also stabilize the gold segregated structures. On
the other hand, in the silver-rich system, difficulty in interpreting the experimental results can be raised because oxidation seems to be present and to provoke the apparent silver segregation. Similarly to our work, the second study also used EDS to characterize the chemical ordering\cite{Martinez2012Jul} and reported Au surface segregation. However, the results were obtained only with one nanoparticle that exhibited equal amounts of gold and silver ($Ag_{0.51}Au_{0.49}$). 

In the present study,  while we consistently observed gold segregation, we did not manage to generate silver-rich nanoparticles using our experimental approach and it is therefore difficult to be assertive that gold segregation remains in the whole composition range of the alloy. However, our machine-learning assisted simulations predict that Au would segregate even for Ag rich compositions. While further experimental studies are required to confirm these trends, we posit that these predictions can be rationalized by combining a literature review of the numerical simulations along with additional calculations, as discussed below. Indeed, the gold segregation at the surface, even for silver-rich nanoparticles, might appear counterintuitive at first sight because it is not found in all experiments, and because it is in contradiction with the surface energy hierarchy ($\gamma$(Ag) is slightly lower than $\gamma$(Au)) and with the atomic size (Ag is slightly bigger than Au). Furthermore, silver surface segregation was consistently found in simulations when using empirical force fields\,\cite{He2018Jan,Du2019Aug,Deng2011Jun,Paz-Borbon2008Apr,Moreira2023Feb}. However, in this context, the simplicity of the employed empirical force fields when compared to our MLIP may lead to inaccurate modeling. In particular, in the work of Paz-Borbon et al.\cite{Paz-Borbon2008Apr}, the Gupta potential seemed at first to demonstrate silver segregation but when further optimization was made at the DFT level, the authors found that gold segregation is favored instead. Similarly, we tested an embedded-atom model (EAM) which was used in the recent paper of Moreira et al.\cite{Moreira2023Feb} and demonstrated that it leads to much more energetic nanoparticles when exhibiting silver surface segregation [See SI\,J].

\begin{figure}[ht]
    \includegraphics[width = 1\linewidth]{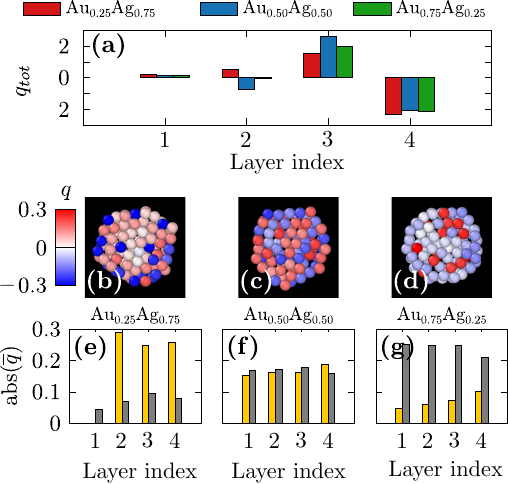}
    \caption{(a) Total Bader charge distribution averaged over the three nanoparticles per chemical composition. (b-d) Slices of typical nanoparticles where atoms are colored following the Bader charges. (e-g) Absolute value of the Bader charge distribution per atomic species averaged over three nanoparticles per chemical composition. We note that the charges are always negative for gold and positive for silver and that only the nanoparticles made of 250\,atoms were studied because of the large computational costs associated with DFT calculations.}
    \label{Fig:charges}	 
\end{figure}

Meanwhile, all of the DFT studies also show the gold segregation both for extended surfaces or for nanoparticles up to few hundreds of atoms.\cite{Cerbelaud2011May,Chen2008Jun,Rapetti2019Mar,Wang2021Dec,Paz-Borbon2008Apr,Gould2014Sep,Hoppe2017Dec} As an explanation, it has been shown from electronic structure investigations that the segregation of the Au atoms at the surface, regardless of composition, is caused by electrostatic forces rather than surface energy. In order to confirm this hypothesis, we performed single-point DFT calculations initialized with the 250-atom nanoparticles found with our freezing simulations and computed the Bader charges. In Figure \protect\ref{Fig:charges}(a), the total charge on the different layers from the center of the NP (index 1) to the surface of the NP (index 4) are shown for the three compositions. It is interesting to note that the surface is always negatively charged and the subsurface is positively charged, by the same magnitude in all the studied chemical compositions. The charge distribution on the different atom types is depicted in Fig.\,\ref{Fig:charges}(b,c,d) as color maps and in Fig.\,\ref{Fig:charges}(e,f,g) as histograms of averaged values taken over atoms in the different layers.  A significant charge transfer from Ag atoms to Au atoms is thus observed with Au and Ag atoms being always respectively negatively and positively charged. A striking observation is that the charge transfer is more observed for the silver rich nanoparticles. Indeed, in the Au$_{0.25}$Ag$_{0.75}$ nanoparticles (Fig.\,\ref{Fig:charges}(b) and (e)), a small number of surface Au atoms bear a negative charge of $\approx$ -0.3\,e whereas the Au atoms at the Au$_{0.75}$Ag$_{0.25}$ nanoparticles surface carry a much lower average charge of $\approx$ -0.1\,e (Fig. \protect\ref{Fig:charges}(d) and (g)). Overall, the total charge of the surface layer remains the same for all of the studied compositions but it is carried by a few but highly charged Au atoms in Au$_{0.25}$Ag$_{0.75}$ nanoparticles and by many but less charged Au atoms in Au$_{0.75}$Ag$_{0.25}$ nanoparticles. These results show that the same effects observed in previous DFT studies on small clusters are also present in larger nanoparticles obtained in our own machine-learning based simulations. They also show that charge transfer between Ag and Au always yields the same surface charge irrespective of the alloy composition. We therefore hypothesize that, as Au atoms become negatively charged when alloyed with silver, they will tend to move towards the surface due to Coulomb repulsion. Further, because the nanoparticle's surfaces then become negatively charged, positively charged Ag atoms are attracted to the subsurface. As the charge transfer from Ag atoms to Au atoms is more pronounced for Ag-rich compositions, the segregation of gold at the surface will be favored even for such compositions, over that of the majority species, as could be intuitively expected. We note that, even if it did not explicitly taking into account the atomic charges, the DNP model as it was fitted to DFT calculations was still able to remarkably translate such complex charge transfer mechanisms into corresponding changes in forces.

\section{Conclusion}

The main goal of this work was to study the formation of gold/silver nanoalloys. Our results in terms of experimental synthesis in the gas phase produced two remarkable observations. First, we found five-fold symmetric particles, including both icosahedron and decahedron with nanoparticles of diameters up to 10\,nm. Second, while surface oxidation can induce silver segregation, only gold segregation was observed in our synthesis results that were obtained in vacuum conditions. These experimental findings were first confirmed by machine-learning assisted simulations. Then, we further explored the chemical phase space by reaching different chemical compositions and confirming the gold segregation and the stability of the five-fold symmetric morphology, even in silver-rich systems. We emphasize that contrary to previous works using classical interaction potentials, the use of MLIP to provide quantum accurate modeling  was key to reproduce the gold surface segregation. An additional advantage of our simulation approach was that we managed to explore the crystallization dynamics that required large-scale simulations with unprecedented chemical accuracy and found that the gold surface segregation occurs before crystal ordering and leads to better crystallization in silver-rich composition.  


By showing that gold surface segregation can be observed in vacuum while silver surface segregation is usually found in more reactive conditions, our study highlights the importance of environmental effects on the chemical distribution of species in a bi-metallic nanoparticle. It also demonstrates the need to study nanoparticles using advanced experimental observations to be able to harness these effects. Further, our results show the importance of taking into account electronic structure effects in nanoalloys, which are impossible to reproduce with conventional empirical potentials. To this end, we demonstrate the tremendous power of MLIP-type potentials for studies of this kind, making it possible to combine the modeling of nucleation processes in realistically-sized systems with DFT precision calculations.



\section*{Acknowledgment}

This study was supported by the French National Research Agency (ANR) in the framework of its “Jeunes chercheuses et jeunes chercheurs” program, ANR-21-CE09-0006 and by the EUR grant NanoX n° ANR-17-EURE-0009 in the framework of the "Programme des Investissements d’Avenir". Computational resources have been provided by CALMIP, by Jean Zay and by TGCC. WAS and CA acknowledge support through  the U.S. National Science Foundation (Award No. CSSI-2003808).  Computational support was provided in part by the University of Pittsburgh Center for Research Computing through the resources provided on the H2P cluster, which is supported by NSF (Award No. OAC-2117681).

\end{document}